\documentclass[nobalancelastpage
,prb
,nofootinbib
,twocolumn
,secnumarabic%,tightenlines%
,amssymb, amsmath,
superscriptaddress,
floatfix]{revtex4-1}

\usepackage{graphicx}% Include figure files
\usepackage{dcolumn}% Align table columns on decimal point
\usepackage{bm}% bold math
\usepackage[T1]{fontenc}
\usepackage[latin1]{inputenc}
\usepackage{latexsym}
\usepackage{amsfonts}
\usepackage{amsmath}
\usepackage{tabularx}
\usepackage{multirow}
\usepackage{fancyhdr}
\usepackage{epsfig}
\usepackage{rotating}
\newcommand{\ud}{\mathrm d}
\newcommand{\degr}{$\,^{\circ}$}
\newcommand{\omob}{$\omega_L/\omega_B\,$}
\begin{document}

%\preprint{AIP/123-QED}

\title{Wide Aperture Vector magnet for neutron scattering studies}% Force line breaks with \\

\author{P. Lavie}
\affiliation{Laboratoire L\'eon Brillouin (UMR 012 CEA/CNRS), IRAMIS, CEA Saclay, b\^{a}timent 563, 91191 Gif sur Yvette, France}
\author{A.M. Bataille}%
 \email{alexandre.bataille@cea.fr}
\affiliation{Laboratoire L\'eon Brillouin (UMR 012 CEA/CNRS), IRAMIS, CEA Saclay, b\^{a}timent 563, 91191 Gif sur Yvette, France}
\author{A. Peugeot}
\affiliation{Service des Acc\'el\'erateurs, de Cryog\'enie et de Magn\'etisme, IRFU, CEA Saclay, b\^{a}timent 125, 91191 Gif sur Yvette, France}%
\author{P. Bredy}
\affiliation{Service des Acc\'el\'erateurs, de Cryog\'enie et de Magn\'etisme, IRFU, CEA Saclay, b\^{a}timent 125, 91191 Gif sur Yvette, France}%
\author{C. Berriaud}
\affiliation{Service des Acc\'el\'erateurs, de Cryog\'enie et de Magn\'etisme, IRFU, CEA Saclay, b\^{a}timent 125, 91191 Gif sur Yvette, France}%
\author{A. Da\"{e}l}
\affiliation{Service des Acc\'el\'erateurs, de Cryog\'enie et de Magn\'etisme, IRFU, CEA Saclay, b\^{a}timent 125, 91191 Gif sur Yvette, France}%
\author{J.-M. Rifflet}
\affiliation{Service des Acc\'el\'erateurs, de Cryog\'enie et de Magn\'etisme, IRFU, CEA Saclay, b\^{a}timent 125, 91191 Gif sur Yvette, France}%
\author{S. Klimko}
\affiliation{Laboratoire L\'eon Brillouin (UMR 012 CEA/CNRS), IRAMIS, CEA Saclay, b\^{a}timent 563, 91191 Gif sur Yvette, France}
\author{J.-L. Meuriot}
\affiliation{Laboratoire L\'eon Brillouin (UMR 012 CEA/CNRS), IRAMIS, CEA Saclay, b\^{a}timent 563, 91191 Gif sur Yvette, France}
\author{T. Robillard}
\affiliation{Laboratoire L\'eon Brillouin (UMR 012 CEA/CNRS), IRAMIS, CEA Saclay, b\^{a}timent 563, 91191 Gif sur Yvette, France}
\author{G. Aubert}
\affiliation{Service des Acc\'el\'erateurs, de Cryog\'enie et de Magn\'etisme, IRFU, CEA Saclay, b\^{a}timent 125, 91191 Gif sur Yvette, France}%

\begin{abstract}
We propose an innovative design for a vector magnet compatible with neutron scattering experiments. This would vastly expand the range of experimental possibilities since applying a magnetic field and orienting the sample in diffraction conditions will become completely independent.
This Wide Aperture VEctor magnet is a setup made of 16 coils, all with a vertical axis. The vertical component of the field is produced by two pairs of coaxial coils carrying opposite currents for an active shielding of the stray field, while the horizontal components are generated by 3 sets of 4 coils each, two above and two below the diffraction plane. This innovative geometry allows a very wide aperture (220$\,^{\circ}$ horizontal, $\pm$ 10$\,^{\circ}$ vertical), which is crucial for neutron diffraction and inelastic neutron scattering experiments. Moreover, the homogeneity of the field is far better than in the usual vertical coils, and the diameter of the sample bore is unusually large (10 cm).
   	The concept has been developed so as to be used as a sample environment on every LLB instrument relevant for magnetism studies.

\end{abstract}

%\pacs{07.55.Db,84.71.Ba,61.05.F-,75.35.-j,75.30.Ds,75.30.Gw,75.50.Ee,75.50.Xx,75.70.-i}% PACS, the Physics and Astronomy
%\keywords{Superconducting magnet, Neutron scattering, Polarized Neutron diffraction, Antiferromagnets, Molecular magnets, Epitaxial thin films}
%Use showkeys class option if keyword
\maketitle

\section{Introduction}

Neutron scattering is a very powerful tool to study condensed matter physics. The general principle of a neutron scattering experiment is to send an intense neutron beam onto a sample and observe in which directions neutrons come out. If the neutron wavelength (given by the de Broglie equation) is of the same order as the lattice parameters of a crystal, neutron diffraction (ND) becomes possible following Bragg's law. The interaction between the neutron spin and the spins of unpaired electrons is rather strong, so magnetic diffraction can be readily performed in order to reveal the magnetic ordering within solids, regardless of the complexity of such order\cite{book_chatterji}. Thanks to recent instrumental progress\cite{filtrage_Gauss3D}, measurements on small samples (like e.g. epitaxial thin films) have recently become much easier. In particular, the systematic study of antiferromagnetic epitaxial thin films is now possible\cite{tricouches_Cr,APL_aimantation_surf,these_Max}.

Moreover, the energy of the so called cold and thermal neutrons used for diffraction is also of the order than the energy of excitations of solids, such as phonons and magnons. Inelastic neutron scattering (INS) measurements then provide direct information about these excitations. This kind of studies is now essentially performed on bulk samples, but will become much easier on thin films with the development of the new generation of time of flight machines. Finally, by using longer wavelength and building instruments dedicated to small angle or reflectivity measurements, it is possible to study larger scale structures, with characteristic lengths in the tens of nm range, which is the realm of nanomagnetism.
In order to widen the range of possible experiments, the Laboratoire L\'eon Brillouin (LLB) wishes to develop a vector magnet with a wide aperture, which would be used as a sample environment on all LLB instruments relevant for magnetism studies.

The paper is organised as follows: section \ref{sec:problematic} describes the problematic leading us to the design of a vector magnet, first by mentioning the existing measurement geometries (sec. \ref{sec:existing_geometries}) and by stating the practical interests of a vector magnet for neutron scattering studies (sec. \ref{sec:practical_interests}). Section \ref{sec:magn_concept} describes the magnetic concept we have designed: we explain the rationale leading to the original design (sec.\ref{sec:config}), then describe the shielding used to limit the stray field (sec. \ref{sec:stray_field})then display the analytical calculations we performed to optimise the field homogeneity (sec. \ref{sec:homogeneity}), and discuss the compatibility of the magnet with polarized neutron scattering experiments (sec. \ref{sec:polarized}).  Finally, section \ref{sec:tech_desc} provides a technical description of the magnet: cold mass (sec. \ref{sec:cold_mass}), cryostat (sec. \ref{sec:cryostat}) and electrical circuit (sec. \ref{sec:electrical_circuit}).

\section{problematic: neutron scattering under magnetic field}
\label{sec:problematic}
\subsection{Existing geometries}
\label{sec:existing_geometries}

\begin{figure}
	\centering
	\centering \epsfig{file=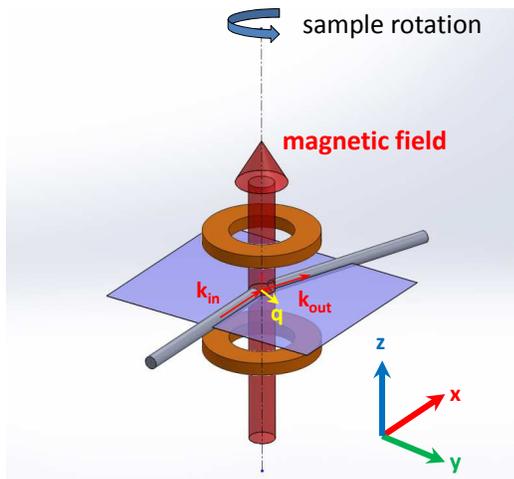,width=.8\linewidth}
	\caption{Neutron diffraction under magnetic field, vertical geometry. The red arrow corresponds to the magnetic field, and $\mathbf{k_{in}}$ and $\mathbf{k_{out}}$ corresponds to the wave vector of the incident and diffracted beam, respectively. The scattering vector is designated by $\mathbf q$. The $x$, $y$ and $z$ directions indicated on the figure give the convention used in the remainder of this paper.}
\label{fig:geom1}
\end{figure}

Given the high sensitivity of neutron scattering measurements to magnetic ordering, magnetic excitations and magnetic domains, the possibility to apply a magnetic field is a rather standard requirement for sample environments used on neutron instruments. While the measurement geometry does not put too stringent constraints on the magnet design in the case of small angle scattering and reflectivity measurements, this is no longer true for diffraction or inelastic measurements: in this case the diffraction angle is rather large (typically from 10 to 120\degr), and in practical designs the objective is to let the horizontal plane containing the incident beam (called the diffraction plane hereafter) as free of geometrical obstacle as possible. Two distinct geometries are widely used:
\begin{itemize}
  \item Vertical field (see FIG. 1) : the sample is mounted on a vertical sample rod, and can rotate around a vertical axis. In this case, the field direction is set by the sample mounting (the field can be rather large, up to 10 T at LLB), and the diffraction plane is almost free of obstacles (typically three sectors, of 20\degr each, are off limits). The part of the reciprocal space accessible during the measurement is thus determined by the direction along which the field should be applied, and is not necessarily optimal for the diffraction or INS measurements.

  \item Horizontal field (see FIG. 2): a set of coils (usually electromagnets) are used to apply the field in a horizontal direction, the sample rotating around a vertical axis. This geometry adds a lot of constraints since large sectors of the diffraction plane are obstructed. If the magnet has a rotation around a vertical axis independent of the rotation of the sample, the field can theoretically be placed along any direction of the plane, put the strong geometrical constraints are often incompatible with setting the sample in diffraction conditions.

\end{itemize}

\begin{figure}
	\centering
	\centering \epsfig{file=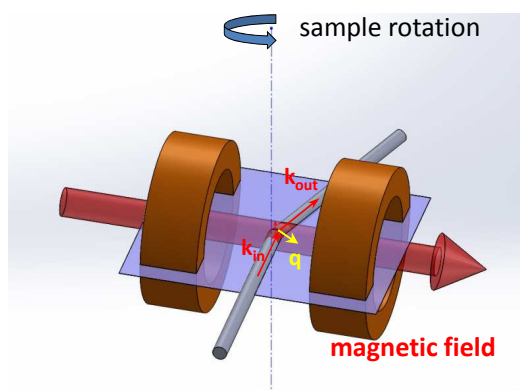,width=.8\linewidth}
	\caption{Neutron diffraction under magnetic field, horizontal geometry. The notations are the same as in figure \ref{fig:geom1}.}
\label{fig:geom2}
\end{figure}

Another type of magnet might be used for neutron scattering experiments: spherical polarimetry experiments require a small guide field to place the spins of the incident neutrons along particular directions. The guide fields need not to be strong (a few mT is usually sufficient to keep the neutron beam polarization) so these small vector magnets are usually made of a set of Helmholtz coils, without soft magnet core. These devices are thus limited to near zero field measurements and restrict the parts of the diffraction plane that are available.

% parler aussi de celui de ISIS
\subsection{Practical interests of a vector magnet}
\label{sec:practical_interests}

The practical benefits of a wide aperture vector magnet can be listed as follows

\begin{itemize}
  \item Precise application of the magnetic field: with the systems now in use, it is difficult to obtain a misalignment of less than a few degrees between the actual field direction and the crystallographic direction along which it should be applied. This misalignment would become vanishingly small with a vector magnet, which would allow measurements on sample with large magnetic anisotropy. This is particularly important in the emerging field of antiferromagnetic spintronics, where anisotropy phenomena play a key role\cite{nmat_park,nmat_marti}
  \item 	Field cooling of samples: by applying a field along a carefully chosen direction of the sample under investigation during its cooling below its magnetic transition temperature, it is possible to obtain a single-domain sample. A vector magnet would thus allow to overcome the detrimental effects of domains\cite{JPCC_Lenertz}, which are twofold: it makes the analysis of the data much more complex (to the point of making it impossible in many cases) and, if the domains are smaller than the coherence length of the beam, it broadens the magnetic peaks (since the area of the peak is constant, this in turn lower the peak intensity, which can then be lost in the measurement noise)
  \item 	Optimization of the dataset: using a 1D vertical field requires choosing the sample orientation with respect to the magnetic field only, as explained above. The part of the reciprocal space accessible during the measurement (which is a cylinder which height is given by the vertical aperture of the coils) may not be the best choice in terms of diffraction (number of Bragg peaks accessible) or inelastic scattering (diffraction plane available) measurements if other sample orientations provide more information. This restriction does not exist with a vector magnet.
  \item 	Local anisotropy measurements: this type of measurement\cite{JPCM_gukasov_brown} are already performed at LLB\cite{PRL_Cao_local_susceptibility}, but require putting the sample out of the magnet, and gluing it again in a new orientation. This procedure is time consuming (typically one or two days of beam time lost for each supplementary orientation) but can be even more detrimental in the case of fragile samples\cite{AC_Ridier,JACS_Ridier}, which may not survive a warming up. A vector magnet would make such experiments much easier and widen the class of appropriate samples to the fragile ones.
  \item 	Polarized neutrons experiments under moderate fields: there is now a gap in the magnetic fields usable for experiments using polarized neutron beams at neutron scattering facilities. In this case, a magnetic field is required on all the path of the incident and (when relevant) scattered beam to keep the beam polarization. This guiding field needs not to be large (a few mT) but it has to vary smoothly. This can be done either by using small vector magnets (for near-zero field experiments) or by using the stray field of large cryomagnets. In this case the field applied to the sample has to be large enough to produce a sufficient stray field, the lower bound being typically 0.5 T. A vector magnet would close this gap (see sec. \ref{sec:polarized} for details).

\end{itemize}

\section{Magnetic concept}
\label{sec:magn_concept}
\subsection{Configuration}

Vector magnets tend to be a combination of the configurations described in FIGS. \ref{fig:geom1} and \ref{fig:geom2}, the vertical component of the field being sometimes produced by a single solenoid placed above the sample. This is not well adapted to neutron scattering experiments, since the diffraction plane is occupied by two sets of coils, making the situation worse than the one depicted on figure \ref{fig:geom2}.

\begin{figure}
	\centering \epsfig{file=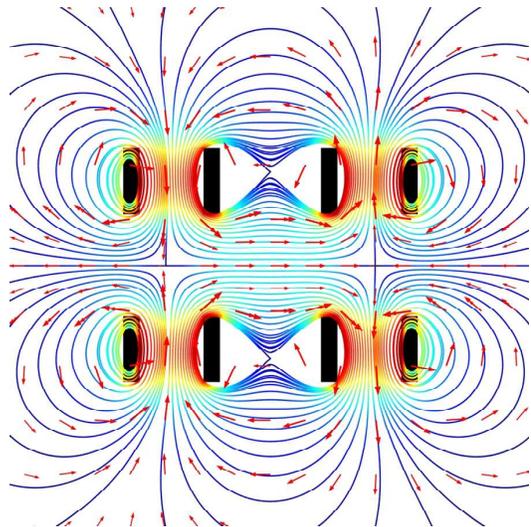,width=.8\linewidth}
	\caption{Field lines produced by a set of four coils with vertical axes, in series/opposition so as to generate a purely horizontal field at the center of the coil quartet. The map represents a vertical cross section of the coil quartet, with the diffraction plane in the middle. The coils are represented by the black rectangles. The color of the field lines are proportional to the field intensity, and the arrows indicates the field direction.}
\label{fig:field_lines}
\end{figure}

The main idea behind the WAVE magnet concept\cite{brevet_US} is to use sets of superconducting coils with \textit{vertical} axes to generate the \textit{horizontal} components of the magnetic field. More precisely, consider sets of 4 coils, two above and two below the diffraction plane. If the currents amplitude are equal in the four coils, and their signs are set as shown on FIG. \ref{fig:field_lines}, the stray field on the center zone of the quartet of coils is horizontal, so it can be used to produce the $B_x$ and $B_y$ components of the field. A possibility would thus be to use a set of 4$K$ independent coils to generate a magnetic field in an arbitrary direction. In this simple configuration, each of the 4$K$ would have to be provided with its own power supply capable of delivering a current in the range $\pm I_0$, which would make the design more complex and much more expansive.

We chose a so called \textit{hybrid} configuration to decrease significantly the number of power supplies. In the hybrid solution, the 4$K$ coils (noted $B_{xy}$ coils hereafter are used for the horizontal components of the generated field, so it is sufficient to have $K$ power supplies which will be connected to those parts of coils in series-opposition of four. Another set of two pairs of coaxial coils (noted the $B_z$  coils in the following) is used to generate the vertical component of the field (hence the hybrid configuration). Each pair is symmetric with respect the $xOy$  plane and their currents flow in opposite senses around $Oz$  for actively shielding the stray field.

\begin{figure}
	\centering
	\centering \epsfig{file=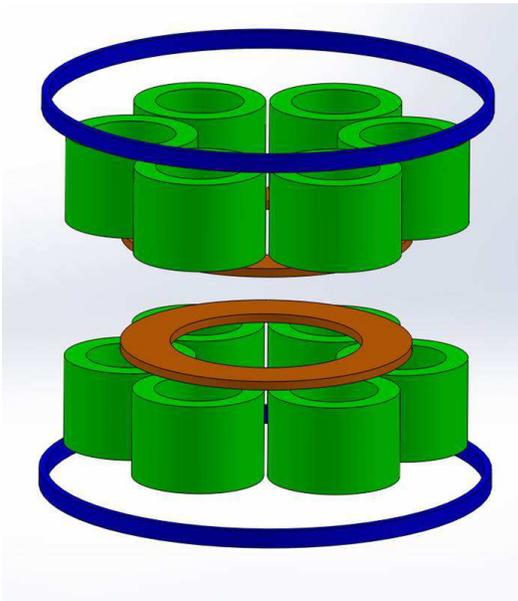,width=.8\linewidth}
	\caption{Possible configuration for the WAVE magnet. The 3 quartets of coils used to generate the horizontal field are represented in green, while the Helmholtz coi pair is represented in orange. The blue coils are used for active shielding (see sec. \ref{sec:stray_field}).}
\label{fig:config}
\end{figure}

A possible layout for the WAVE magnet is thus given on FIG \ref{fig:config}. In principle, two quartets of coils are sufficient to generate the horizontal field components but we explored solutions with $K$ quartets of coils, and found that $K=3$ configurations have significant advantages in terms of field homogeneity, angular acceptance and compatibility with polarized neutron measurements. In this configuration, only 4 power supplies are needed.

% traduire une partie de la note de Guy Aubert
\label{sec:config}
\subsection{Stray field and shielding}
\label{sec:stray_field}

Stray field generated by the device has to be taken into account at the early stages of the design, since it can be troublesome for the electrical motors included or close to the device. The motor used in standard neutron diffractometers can withstand rather significant fields, but this is not the case of cryocoolers required to cool the sample to be investigated or, in our case, the superconducting coils. The WAVE magnet is designed so that the field at the cryocoolers positions never exceeds 3 mT. This is not a stringent requirement for the $B_{xy}$ coils, given their mounting in series/opposition, and no specific precaution is needed in this case. This is no longer true for the $B_z$ coils. In this case we use {\it active shielding}, namely a second set of coils with the same axis and center as the first, dimensioned so that the total dipole moment of both sets of coils cancels out together with the second degree term of the central field harmonic expansion which achieves the required field homogeneity in the sample zone (see \ref{sec:hom_vert_field}).  This does not require another power supply since both coils are mounted in series.

\subsection{Field homogeneity}
\label{sec:homogeneity}
A specific aspect of neutron scattering measurements is that the sample can be rather large (up to 10 mm long, sometimes even more), and that the neutron beam is also large (typically 20$\times$10 mm). The field homogeneity requirements are thus more stringent than for techniques using smaller beams such as x-ray based techniques (the vertical size of a synchrotron-generated beam is at most 100 $\mu$m, and it can be focused to much smaller sizes) or magneto-optical Kerr effect (the laser beam diameter is about 1 mm, much less for MOKE imaging).

\subsubsection{Analytical protocol}

Optimization of the field homogeneity was obtained through analytical calculations, of which we will give the main ingredients here. We use the scalar pseudo potential $V^\star$ as a calculus intermediary to obtain the magnetic field using the relation:

\begin{equation}\label{eq:pseudo_V}
  \vec B = \mu_0\overrightarrow{\mathrm{grad}}V^\star
\end{equation}

Since $V^\star$ is a solution of the Laplace equation, it can be expressed in spherical coordinates using a fairly general equation:

\begin{equation}\label{eq:spher_harm}
  V^\star=\frac{1}{\mu_0}\sum_{l=1}^{+\infty}\sum_{m=-l}^l \left(\frac{\alpha_l^m}{r^{l+1}}+\beta_l^m r^l\right)Y_l^m(\theta,\varphi)
\end{equation}
with $\alpha_l^m$ and $\beta_l^m$ being constants, and $Y_l^m$ corresponding to the spherical harmonics. In our case, this development will be used at or near the origin, so all the $\alpha_l^m$ shall be zero since $V^\star$ is finite. We can thus write:
\begin{eqnarray}\label{eq:def_Vstar}
  V^\star(r,\theta,\varphi)=\sum_{l=0}^{+\infty}\sum_{m=0}^lV_l^mr^lY_l^m(\theta,\varphi)\\
  \label{eq:def_B}
B(r,\theta,\varphi)=\sum_{l=0}^{+\infty}\sum_{m=0}^lB_l^mr^lY_l^m(\theta,\varphi)
\end{eqnarray}

Equation \ref{eq:pseudo_V} allows us to easily give the relation between the $V_l^m$ and the $B_l^m$

\begin{equation}\label{eq:rel_Vlm-Blm}
  B_l^m=\frac 12\left(V_{l+1}^{m-1}-(l+m+1)(l+m+2)V_{l+1}^{m+1}\right)
\end{equation}

\subsubsection{Coils for vertical field components}
\label{sec:hom_vert_field}

Each coil is an $Oz$  axis toroid of rectangular section (inner radius $a_1$, outer radius $a_2$, lower axial end $b_1$ , upper axial end $b_2$ ) carrying a uniform azimuthal current density $j_0$. A first pair of coils $(a_1,a_2,b_1,b_2,j_0)$ and $(a_1,a_2,-b_2,-b_1,j_0)$ symmetrical with respect the $xOy$  plane generates a central axial field $B_0$ . A second coaxial pair located further from the center $(a_1^\prime,a_2^\prime,b_1^\prime,b_2^\prime,-j_0)$ and $(a_1^\prime,a_2^\prime,-b_2^\prime,-b_1^\prime,-j_0)$ generates a negative central field $B^\prime_0$, with sign opposite to $B_0$. The 8 geometrical parameters and the absolute value of the current density $j_0$ result from a nonlinear optimization process minimizing the volume of the superconducting windings under dimensional constraints and nonlinear magnetic constraints.
The dimensional constraints consists in setting $b_1$  for allowing the required free central space and the outer coils dimensions $a^\prime_2$ and $b^\prime_2$ since the superconducting volume tends to decrease when these dimensions increase, which is quite counter intuitive.
The magnetic constraints are the value of the central field coefficient $(B_0+B_0^\prime)/j_0$ and the cancellation of both the total dipole moment and the second degree coefficient of the central axial field expansion as a function of $z$ , the inhomogeneity being thus governed by the fourth degree one. Analytical expressions are known for these three nonlinear constraints as functions of the 8 geometrical parameters and the 5 unknown ones result from a standard nonlinear constraint optimization process.
The actual design of the coils must then take into account the usual parameters for superconducting coils (critical current, peak field, electromagnetic stresses).

\subsubsection{Coils for horizontal field components}

We consider a set of $N$ pairs of identical coaxial axisymmetric coils the axis of which is parallel to $Oz$ at a distance $a_0$. These $N$ axes are regularly distributed around $Oz$   with azimuths $\varphi_j=2\pi j/N$, with $j=0,1,...,N-1$ . The two coils of a pair are geometrically symmetrical with respect the $xOy$  plane, the upper coil carrying a current intensity $I_j$  and the lower one the opposite $-I_j$.

The first problem is to find the distribution of intensities $I_j$ leading to a total maximum field

   \begin{equation}\label{eq:defBHorz}
     \vec B = \sum_{j=0}^{N-1}\vec B_j=B_0\vec u_0
   \end{equation}

at the center $O$ of the magnet, in the direction $(Ox,\vec u_0)=\varphi_0$  of the $xOy$ plane. This distribution must satisfy the two following equations:

\begin{eqnarray}
  \sum_{j=0}^{N-1}I_j\sin\left(\varphi_0-j \frac{2\pi}{N}\right)&=&0 \\
  \sum_{j=0}^{N-1}I_j\cos\left(\varphi_0-j \frac{2\pi}{N}\right)&=&GB_0
\end{eqnarray}
where $G$  is a geometrical factor characteristic of the geometry of a single coil. Achieving any direction $\varphi_0$ requires $N\geq 3$ but given $\varphi_0$, $G$  and $B_0$  one gets an infinity of solutions among which the "best" one must be chosen. To that purpose we use the p-norm of the vector of intensities:

\begin{equation}\label{eq:def_pnorm}
  \|\vec I\|_p=\left(\sum_{j=0}^{N-1}|I_j|^p\right)^{\frac 1 p}
\end{equation}

Minimizing the 1-norm (which corresponds tu current carrying capacity) would be the best solution for superconducting magnets but the corresponding mathematical analysis is highly complicated and does not lead to an analytical solution of the homogeneity problem. The same difficulty arises for the $\infty$-norm (maximum intensity) while the Euclidean one (the 2-norm) leads to a simple elegant solution:

\begin{equation}\label{eq:expr_Iopt}
  I_j=\frac{2GB_0}N \cos\left(\varphi_0-j \frac{2\pi}{N}\right)
\end{equation}

One can easily verify that this expression gives a value of the 1-norm very close to its true minimum which can only be obtained by a numerical analysis.
Using this simple law for the distribution of intensities, one can get the spherical harmonic expansion (SHE) of the total field in the central empty region from that of its scalar potential. This expansion for the single upper magnet, in the case where $j=0$ and $\varphi_0=0$ reads:

\begin{equation}\label{eq:expr_Vstar}
  \frac{V^\star}{\mu_0I_0}=\sum_{\ell=1}^{+\infty}r^\ell\left[Z_\ell P_\ell(\cos\theta)+\sum_{m=1}^\ell X_\ell^mP_\ell^m(\cos\theta)\cos m\varphi\right]
\end{equation}

where $P_\ell$ are Legendre polynomials of degree $\ell$  and $P_\ell^m$  are associated Legendre functions of degree $\ell$ and order $m$  with $P_1^1(\cos\theta)=\sin\theta$. The coefficients $Z_\ell$  and $X_\ell^m$ depend on $a_0$ and on the geometrical parameters of the coil.

A simple, yet tedious calculation, leads to the conclusion that the only non vanishing $X_\ell^m$ coefficients in eq. \ref{eq:expr_Vstar} ar those for which $m=k\times N\pm 1$
Another symmetry consideration considerably simplifies the SHE: since the electrical current distribution generating the field is antisymmetric with respect to the horizontal plane containing the origin, the field has to be symmetric with respect to that same plane, and the potential antisymmetric. Given the parity properties of the Legendre polynomials, this implies that the only non zero $V_\ell^m$ are those for which $\ell+m$ is odd.

\begin{figure}[h!]
	\centering
        \epsfig{file=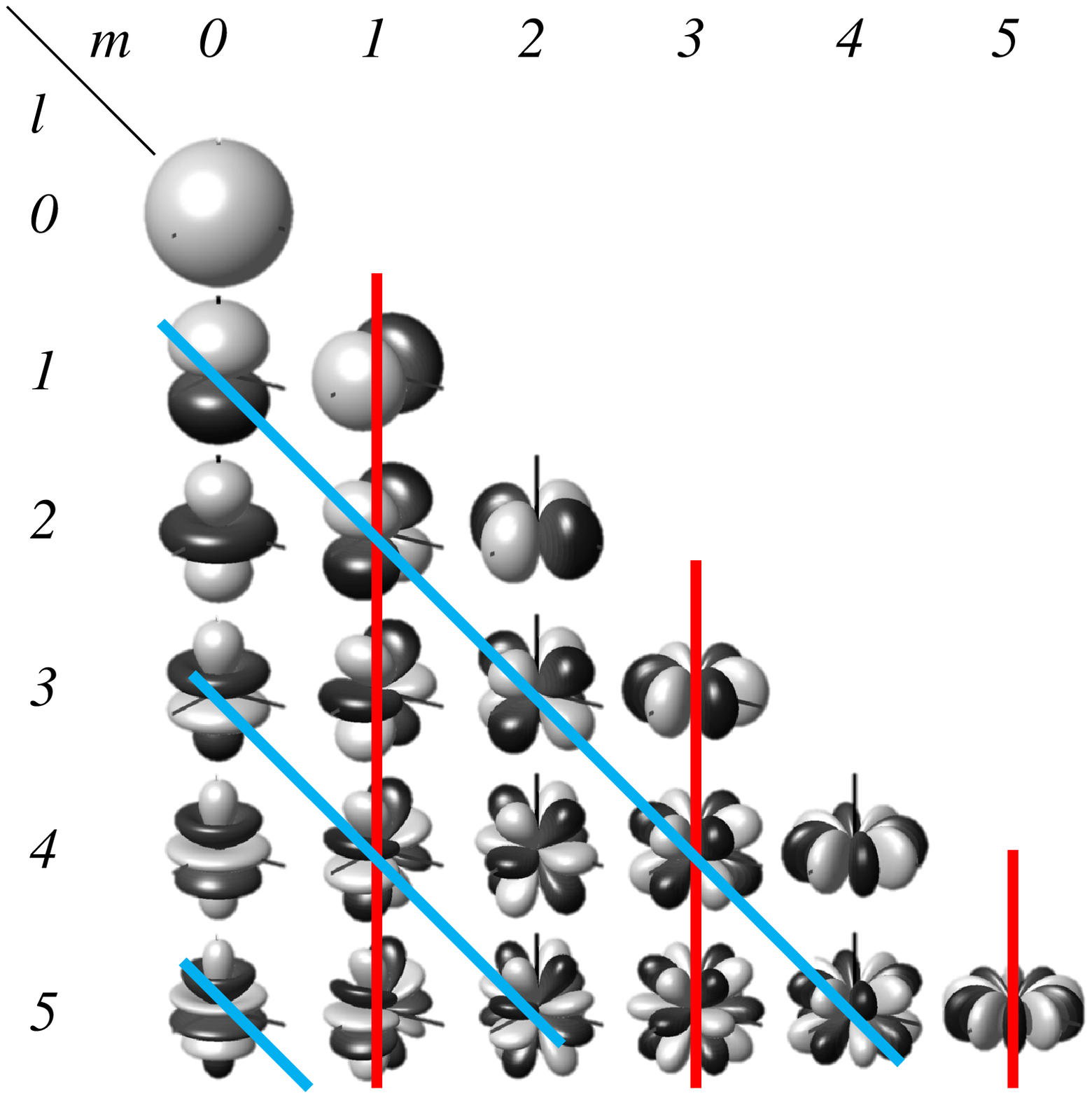,width=.8\linewidth}
	\caption{Schematic representations of the $Y_l^m$ functions, and elimination of terms in the spherical harmonics development of the field by symmetry considerations for $N=2$ or 3. The blue lines corresponds to terms vanishing because $\ell + m$ is odd, while the red lines corresponds to harmonics not fulfilling the $m=kN\pm 1$ condition.}
	\label{fig:harmo_sph_sym}
\end{figure}

The symmetry allowed and forbidden terms for the $N=4$ and $N=6$ cases are summed up on figure \ref{fig:harmo_sph_sym}. The combination of the two criteria eliminates all terms corresponding to odd powers of $r$ in the spherical harmonics development of $B$. Deriving the SHE of the total scalar potential for the $N\geq 3$ pairs of coils is a little tricky and leads to the following expression:
\begin{widetext}
\begin{eqnarray}
  \frac{V^\star}{\mu_0NI_0} &=& \sum_{k=1}^{+\infty}\sum_{n=0}^{+\infty}r^{kN-1+2n}X_{kN-1+2n}^{kN-1}\times\cos[(kN-1)\varphi-\varphi_0]P_{kN-1+2n}^{kN-1}(\cos\theta)\nonumber \\
   &+& \sum_{k=0}^{+\infty}\sum_{n=0}^{+\infty}r^{kN+1+2n}X_{kN+1+2n}^{kN+1}\times\cos[(kN+1)\varphi-\varphi_0]P_{kN+1+2n}^{kN+1}(\cos\theta)
\end{eqnarray}
\end{widetext}

The principal term corresponds to  :

\begin{equation}\label{eq:epxrVstarfin}
  V^\star=\mu_0NI_0X_1^1\sin\theta\cos(\varphi-\varphi_0)
\end{equation}

That corresponds to a constant field $B_0=-\mu_0NI_0X_1^1$  in the direction  $(Ox,\vec u_0)$ of the $xOy$ plane. A careful analysis of the above expression of the scalar potential shows that if $N\geq 5$, the two following terms of its SHE are proportional to the single coefficient $X_3^1$ . It is thus sufficient to choose the geometrical parameters of the individual coil which cancel this term out for having as first "inhomogeneous" terms degree five ones. For instance, in the privileged case $N=6$ , these leading inhomogeneous terms are $X_5^1$ and $X_5^5$.

For an individual toroidal coil of rectangular section $(a_1,a_2,b_1,b_2)$, the axis of which is at a distance $a_0$  of $Oz$ , one can get an analytical expression of $X_3^1$ as a function of these five parameters and one has simply to minimize the volume of superconductor subject to the nonlinear constraints of the value of the field coefficient and $X_3^1$.
$N>6$ configurations can also be discarded by considerations not related to homogeneity, such as the number of power supplies required (and hence the cost of the magnet) and the diameter of the sample bore.

\subsection{Compatibility with polarized neutron scattering experiments}
% mettre ici les calculs matlab fait il y a un an
\label{sec:polarized}
\subsubsection{General principle}
One of the major uses of neutron scattering is to study magnetism, since the interaction between the neutron spin and that of unpaired electrons is rather large. Unpolarized neutron beams can be used to study magnetic ordering as soon as the magnetic period is different from the structural one, but this criterion is not always met. In particular, studies of paramagnets such as molecular magnet crystals\cite{AC_Ridier,JACS_Ridier} require the use of a polarized neutron beam. The ability to keep the beam polarization is thus a prerequisite of any magnet used in a neutron scattering facility, including the WAVE vector magnet. The mathematical criterion comes from the comparison between two frequencies:

\begin{equation}\label{eq:Larmor}
  \omega_L=|\gamma_N|\|B\|
\end{equation}

is the Larmor frequency ($\gamma_N\simeq$ 1.83 10$^8$T.s$^{-1}$) corresponding to neutron precession and

\begin{equation}\label{eq:omegaB}
  \omega_B=\left\|\frac{\ud\vec{u}_B}{\ud t}\right\|
\end{equation}

is the frequency associated with the variation of the magnetic field in the neutron reference frame ($\vec u_B$ is the unit vector defining the field direction). The neutron beam polarization is conserved as long as $\omega_L\gg\omega_B$ (or $\omega_L\ll\omega_B$ for a very short time, provoking what is called a non adioabatic transition between parallel and antiparallel alignments of the neutron spin with respect to the field). Practically speaking, polarization loss happens in two cases:
\begin{itemize}
    \item when the neutron beam goes across an extended small field zone, $\omega_L$ becomes too low. A well known example of that is the zero field zone in the medium plane of Helmholtz coils. We shall address this point in the following.
    \item when the spatial variations of the field are fast enough, $\omega_B$ becomes comparable to $\omega_L$. Since the neutrons of the incident and diffracted beams move essentially at a constant speed $\vec v$, the right hand side of equation \ref{eq:omegaB} is in fact a spatial derivative.
\end{itemize}

The geometrical configuration of the WAVE magnet offers a solution to solve this problem: since $K=3$, there is no unique solution to obtain a given magnetic field. If we add a constant intensity to the values defined by eq. \ref{eq:expr_Iopt}, the field at the sample position is unchanged (since the vectorial sum of the supplementary fields is zero), yet the stray field in the horizontal plane may be increased. In the following, we numerically explore the possibilities of this {\it active zero field} procedure in two difficult cases: a purely vertical magnetic field, and a purely horizontal one. To keep the discussion short in this general paper, we limit ourselves to the {\it half polarized} case, where only the conservation of the polarization of the incoming beam is evaluated (a more detailed study on the fully polarized case will come latter).

Given the electrical configuration of each quartet of coils, the common vertical axis to two set of $B_{xy}$ coils is a problematic zone for keeping the neutron beam polarization, since the spatial derivative becomes infinite at the intersection of the axis and the equatorial plane, so the pulsation defined by eq. \ref{eq:omegaB} is large on a wide zone. The incident beam path is thus fixed in the middle of two such axes, where the field spatial variations are smoother.

We choose a rather conservative approach, {\it i.e.} we consider that the beam polarization is conserved provided the \omob ratio is greater than 100 on the whole incident beam path. The beam is a parallelepiped with transverse dimension of 20$\times$20 mm. All the results were evaluated for $\lambda$=1.4 \AA, which is the polarized wavelength of the 6T2 diffractometer and is not a favorable case since it increases $\omega_B$ compared to longer wavelength.

\subsubsection{Vertical magnetic field}

    \begin{figure}[h!]
	\centering
        \epsfig{file=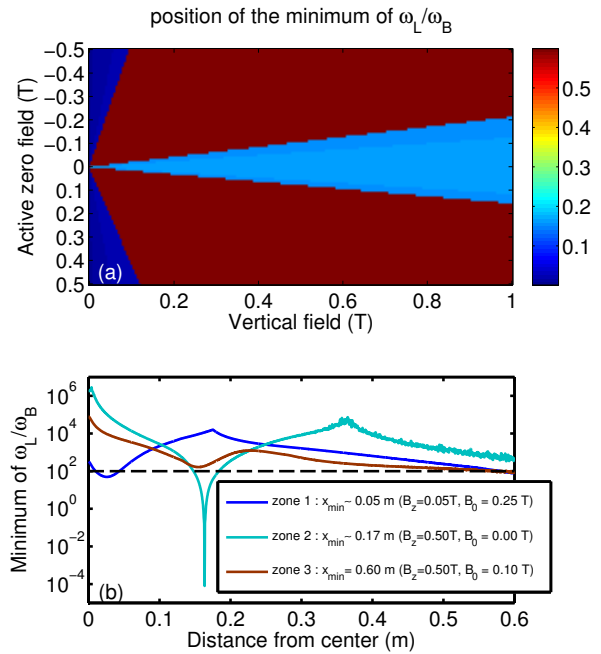,width=.9\linewidth}
	\caption{(a): position of the minimum of the $\omega_L/\omega_B$ ratio. We observe three distinct regimes: Dark blue zone, the worst point is near the sample position. Light blue zone, it is around 170 mm from the sample axis. Brown zone, it is at the very end of the simulated beam, very far from the sample axis and outside the magnet. (b) : examples of the evolution of the \omob ratio as a function of the position along the incident beam. ($x=0$ corresponds to the sample), for each of the three regimes described on the left panel (same color code).}
	\label{fig:worstX_Bvert}
\end{figure}

The case of a purely vertical field is a well known unfavorable case for Helmholtz coils, since they are symmetrical with regards to the diffraction plane. Figure \ref{fig:worstX_Bvert} (a) shows the location of the point where the \omob ratio is smallest. One can identify three regimes:
\begin{itemize}
  \item zone 1: the minimum is close to the sample axis
  \item zone 2: the minimum is roughly at 170 mm from the sample axis. This corresponds to the usual annulment zone of Helmholtz coils
  \item zone 3: the minimum is at the end of the simulated zone
\end{itemize}

\begin{figure}[h!]
	\centering
        \epsfig{file=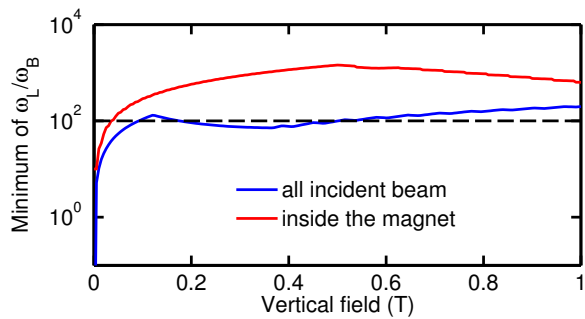,width=.9\linewidth}
	\caption{Minimum of the \omob ratio on the whole incident beam, in the optimal active zero field conditions. The dashed line corresponds to a ratio of 100.}
	\label{fig:bestR_Bvert}

\end{figure}

Figure \ref{fig:worstX_Bvert} (b) gives examples of the evolution of the \omob ratio (more precisely the minimum of \omob at a given position along the incident beam) for each of the three regimes. The minimum is very sharp in the case of zone 2, and \omob is small on a rather large length of beam. This is the well known problem of symmetrical coils for neutron measurements. In the case of zone 1, the value at the minimum is not so small, so a significant polarization (yet far from the initial one) may still be present at the sample position. The most interesting regime in terms of polarized neutron measurements is the third one, since the minimum of \omob is observed outside the magnet, so guiding fields may be used to solve the problem. Figure \ref{fig:bestR_Bvert} gives the lowest value of \omob obtained in the optimal active zero field conditions, as a function of the vertical field. The main effect of the active zero field is to put the minimum of \omob outside the magnet, so that \omob is well above 100 inside the magnet for vertical fields 50 mT or greater. This is a much smaller value of operation than what is now obtained with the standard scalar magnets adapted to neutron scattering studies.

\subsubsection{Horizontal magnetic field}

    \begin{figure}[h!]
	\centering
        \epsfig{file=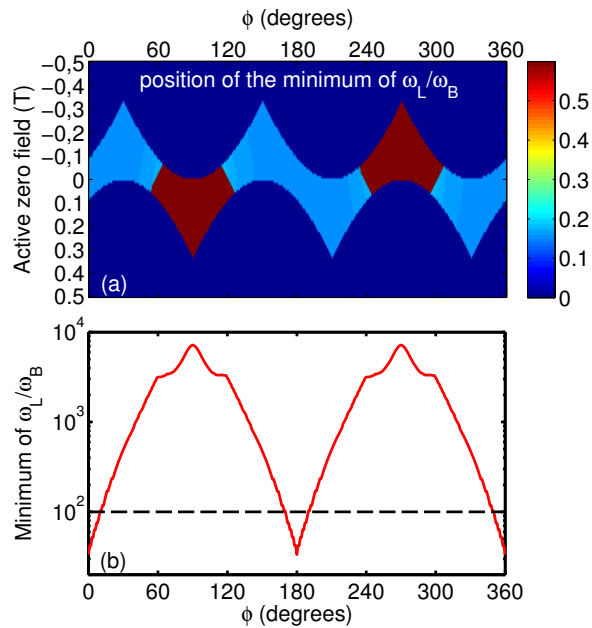,width=.9\linewidth}
	\caption{(a): position of the minimum of the $\omega_L/\omega_B$ ratio. We observe only two distinct regimes: light blue zones, it is around 150 mm from the sample axis. Brown zones, it is at the very end of the simulated beam, very far from the sample axis and outside the magnet. The dark blue zones corresponds to forbidden electrical configurations, in which the current in at least one of the quartet of coils is above the maximum values allowed by the power supplies. (b) Minimum of the \omob ratio on the whole incident beam trajectory, in the optimal active zero field configuration, as a function of the angle between the field and the incident beam direction. The dashed line corresponds to a ratio of 100.}
	\label{fig:worstX_bestR_Bhorz}

\end{figure}

\begin{figure}[h!]
	\centering
        \epsfig{file=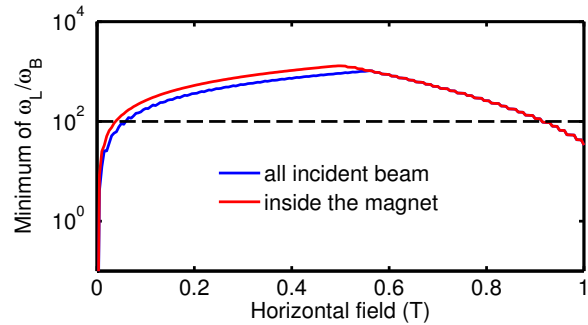,width=.9\linewidth}
	\caption{Minimum of the \omob ratio on the whole incident beam, in the optimal active zero field conditions. The graph corresponds to $\varphi$=0 but could also apply to $\varphi$=180, the curves being identical. The dashed line corresponds to a ratio of 100.}
	\label{fig:bestR_Bhorz_phi0}

\end{figure}

Another configuration for which keeping a significant incident beam polarization is difficult is the case of a purely horizontal field. In this case, the active zero field method may not be as efficient, since the value of the active zero field is restricted by the critical current in the coils and/or the current delivered by a given power supply. This is illustrated on figure \ref{fig:worstX_bestR_Bhorz} (a), which displays the location of the minimum of \omob as a function of the field angle, for a 1 T field. The allowed current distribution makes only a small fraction of the number of cases simulated, since in this case one of the power supplies delivers a current not far from the maximum value. As shown on figure \ref{fig:worstX_bestR_Bhorz} (b), this is not a problem for most of the angles, except when the field direction is close to the incident beam direction. We have simulated this specific case, as illustrated by figure \ref{fig:bestR_Bhorz_phi0}. In this case the minimum of \omob is within the magnet for fields above 0.6 T, and in any case when the minimum is outside the gain is minimal, contrary to the vertical field case. The \omob ratio can be kept above 100 for fields between 0.04 and 0.91 T, again a pretty wide range which corresponds to values not well suited for polarized neutron measurements with the usual magnets. Even at 1 T, \omob is still about 30, which may limit the polarization loss.

\section{Technical description of the WAVE magnet}
\label{sec:tech_desc}
% mettre dans cette partie les aspects "technological innovations du projet inter-DIM
\subsection{Cold Mass}
\label{sec:cold_mass}

\begin{figure}
	\centering \epsfig{file=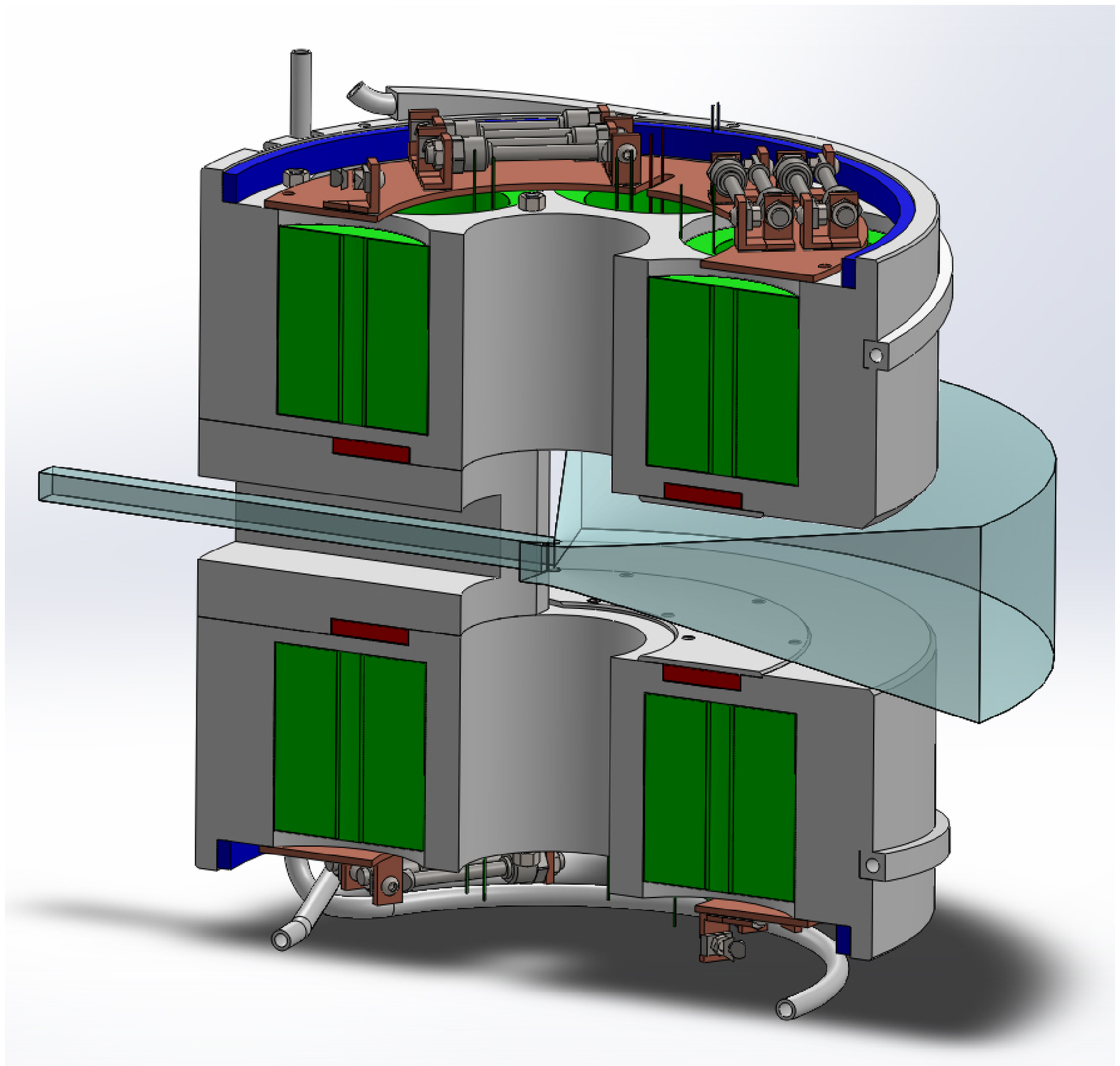,width=.9\linewidth}
	\caption{Vertical cross section of the WAVE magnet, seen in perspective. The incident beam is represented in shaded blue, as the angular aperture available for the scattered beam.  Note the large aperture of the cold mass, 220$^\circ$ horizontally and $\pm$10$^\circ$ vetically.}
\label{fig:vert_cut}
\end{figure}

The cold mass consists in two large aluminum boxes mounted symmetrically with the diffraction plane. A gap is machined in the lower and upper boxes to leave a free space for the beam of 60mm , corresponding to an equatorial vertical opening angle of $\pm$10\degr. The clearance angle in the horizontal plane is 220\degr. Only a sector of 140\degr is kept for mechanical reasons between the two boxes. A channel is machined within this sector for the passing of the incident neutron beam. The aluminum boxes are used as coil casings and as thermal exchangers with a circuit of aluminum pipes fixed on the boxes.

The lower and upper boxes are machined to receive the following windings:
\begin{itemize}
  \item  6 $B_{xy}$ coils (in green on figure \ref{fig:vert_cut}) Their axis is situated on a diameter 300 mm and each coil has the following dimensions: inner radius 49 mm, outer radius 73 mm and height 100 mm.
  \item 1 $B_z$ coil (orange on figure \ref{fig:vert_cut}) on the vertical axis of the experiment .with the following dimensions: inner radius 106 mm , outer radius  161 mm and height 10 mm.
  \item 1 shielding coil (in blue on figure \ref{fig:vert_cut}) with the following dimensions: inner radius 241 mm, outer radius 250 mm and height 19 mm.

\end{itemize}
	
%
%\begin{figure}
%	\centering \epsfig{file=horz_cut.eps,width=.98\linewidth}
%	\caption{Horizontal cross section of the WAVE magnet. The lower part of each coil is represented: $B_{xy}$ coils in green, $B_z$ coil in orange and shielding coil in blue. Note the very large horizontal aperture, set to 220\degr}
%\label{fig:horz_cut}
%\end{figure}

\subsection{Cryostat}
\label{sec:cryostat}

\begin{figure}
	\centering \epsfig{file=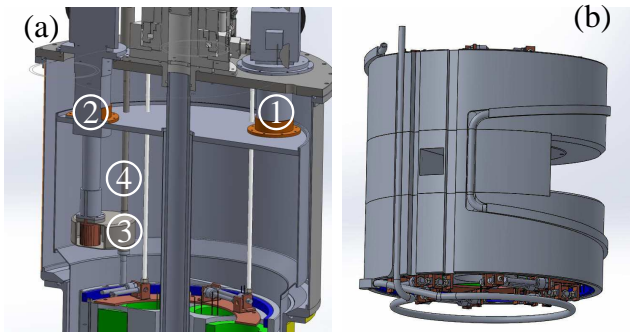,width=.98\linewidth}
	\caption{The cryogenic system of the WAVE magnet. (a) main components of the system. 1 single stage cryogenerator cooling down the upper stage of the cryostat don to 50 K stage. 2 : double stage cryogenerator which cools down the dual phase cryogenic fluid. 3: He reservoir, containing 1 L of liquid He. Note the copper winglets used to liquify the gaseous fraction of He present at the end of the theromsiphon circuit. 4: external He bore, used to fasten the first cooling of the magnet. (b) thermosiphon circuit. Note the first turn at the bottom of the cryostat, required to cool down the electrical buses. Another turn is present at the top.}
\label{fig:cryo}
\end{figure}

The design of the cryostat and of the cooling system(see FIG. \ref{fig:cryo}) is based on two G-M cryocoolers (single stage and two-stage needed for 4.5 K and 50 K heat loads) and a small thermosiphon using aluminum pipes as exchangers along the cold mass at 4.5 K and a helium phase separator of 1 liter fixed on the 2nd stage of the two-stage cryocooler. The helium flow is estimated to 1.5 g.s$^{-1}$, which is consistent with the 1W cooling power of the cryocooler at 4K. If needed, a quick cool-down from 300 K is achieved by using a temporary external supply from LHe Dewar.
The vacuum vessel is made of four main parts: an outer cylinder, an inner cylinder and an annular lower flange. These three elements will be in aluminum to reduce the interaction with the beam of neutrons. The inner cylinder is designed as an \textit{anti-cryostat} and will host the sample and the associated instrumentation that are not part of the project. The fourth part of the cryostat is the upper flange made of stainless steel. This upper flange will have 9 ports: 2 for the G-M cryocoolers , 5 ports for the current leads and 2 for the voltage taps and magnet  instrumentation. Static insulation vacuum is achieved by using a temporary pumping unit on a dedicated lateral port.

The other parts of the cryostat are the thermal shield, also in aluminum, the multilayer insulation, the helium reservoir with the helium pipes of the thermosiphon circuit and the cold-to-warm supports.

\subsection{Electrical circuit}
\label{sec:electrical_circuit}

\begin{figure}[ht!]
	\centering \epsfig{file=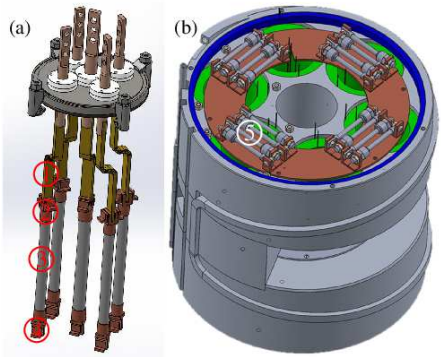,width=.7\linewidth}
	\caption{(a): Current leads. 1 resistive part, made of brass. 2: junction between resistive and superconducting parts, made of copper to allow thermal dilation. 3: high-$T_C$ superconductor part, made of Bi 2212. 4: junction with the low-$T_C$ part, made of copper to allow thermal dilation. (b): Series of diodes used for electrical protection. the number 5 designates a set of 4 such diodes.}
\label{fig:diodes}
\end{figure}

The electrical circuit will be made of 4 independent power supplies able to deliver up to 250 A each. During unpolarized neutron measurements, the values of currents flowing through the coils generating the horizontal components of the field is made according to eq. \ref{eq:expr_Iopt}, which ensures that the total current in the magnet is lower than 450 A. This is no longer true when active zero field is used, yet the optimal conditions are consistent with a total current lower than 500 A. The current leads are thus designed accordingly: 4 independent leads to send current through the 4 sets of coils, but a common lead for collecting it after. The current leads (see FIG. \ref{fig:diodes} (a)) are made of two parts: a resistive one made of brass, which temperature will go from 300 to 50K,  and a superconducting one made of a hollow ceramic bar of high-$T_C$ superconductor (the so called Bi2212). The field is significant only in the lower part of the high-$T_C$ current lead, where the critical field is very large since the temperature is close to 4K. Electrical buses and protection diodes will be mounted on copper rings, in order to facilitate their thermalization. Quench protection will be ensured by 32 diodes (2 for each of the 16 individual coils), the top part being shown on figure \ref{fig:diodes}. In case of quench, the whole current discharge takes less than one second and leads to a warming of 106 K at the hottest point, which is acceptable. The maximum voltage across a quenched coil is estimated to 200 V.

\subsection{Implementation at LLB}
\label{sec:impl_LLB}
\begin{figure}[ht!]
	\centering \epsfig{file=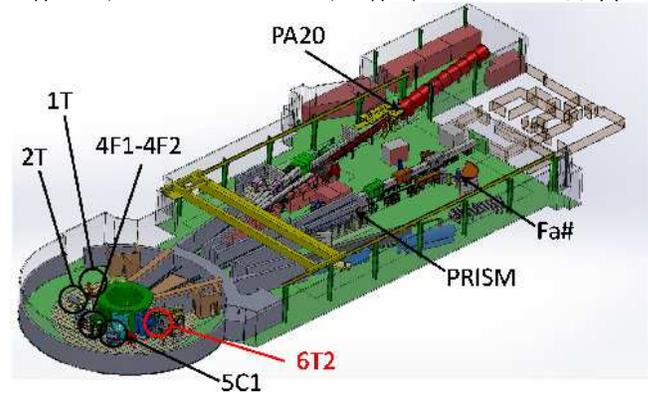,width=.98\linewidth}
	\caption{Possible implementation of the WAVE magnet at LLB. Single-crystal diffractometer: 6T2, 5C1. Triple-axis spectrometers: 1T,2T,4F1,4F2. Polarized Neutron reflectivity: PRISM. Small angle scattering: PA20. Time of flight inelastic scattering: Fa\#}
\label{fig:LLB}
\end{figure}

LLB is a user facility which offers a comprehensive instrumental suite of 22 spectrometers, and new instruments are currently under commissioning or development (see figure\ref{fig:LLB}). The Wave magnet is designed so that it can be used as a sample environment on all the instruments relevant for magnetism studies. This includes:

\begin{itemize}
  \item Diffraction : two instruments are suitable for magnetism studies: 5C1 and 6T2, which both offer polarized neutron beams. The magnet comissioning will be performed on the 6T2 diffractometer, which is the more versatile of the three and puts the most stringent constraints in terms of external diameter of all the instruments on which WAVE could be used.
  \item Inelastic Neutron Scattering: the WAVE magnet shall be used both on the cold (4F1, 4F2) and thermal (1T, 2T) triple axis spectrometers available at LLB. The WAVE magnet could also be used on the future time-of-flight machine, Fa$\sharp$, currently being build.
  \item Small Angle Neutron Scattering and neutron reflectivity : the WAVE magnet will be compatible both with the PRISM polarized neutron reflectometer, and the PA 20 small angle scattering instrument, currently being commissioned.
\end{itemize}

\section*{Conclusions}

We propose an innovative design for a vector magnet compatible with wide angle neutron scattering experiments. The original geometry consists of 16 coils with wertical axes, the horizontal components of the field being generated by the stray fields of sets of 4 coils in series opposition. The proposed design can apply fields up to 1 T in any directions of space, and allow a very large angular aperture (220\degr horizontal, $\pm$10\degr vertical). The homogeneity of the field is very high (better than 50 ppm in a 5 mm radius sphere), and the magnet is compatible with polarized neutron scattering experiments. The magnet is designed so as to be used on all the instruments of LLB which are relevant for magnetism studies, and should thus open a wide range of new possibilities.

\acknowledgements

This work is supported by the Region Ile de France through the NANOK-2014-ML-005 grant.
The authors would like to thank M. Lenertz for his suggestions concerning the manuscript.

\end{document}